\documentstyle[art10]{article}
\begin{document}
\baselineskip=18pt

\title{\bf Parallel Cluster Labeling \\
for\\
 Large-Scale Monte Carlo Simulations}

\bigskip

\author{
M. Flanigan$^*$ and P. Tamayo \\
{\it Thinking Machines Corp.} \\
{\it Cambridge, MA 02142.}
}
\maketitle

\bigskip

\begin{center}
{\bf ABSTRACT}
\end{center}

\noindent

 We present an optimized version of a cluster labeling algorithm previously
introduced by the authors.  This algorithm is well suited for large-scale
Monte Carlo simulations of spin models using cluster dynamics on parallel
computers with large numbers of processors.  The algorithm divides physical
space into rectangular cells which are assigned to processors and combines
a serial local labeling procedure with a {\it relaxation} process across
nearest-neighbor processors.  By controlling overhead and reducing
inter-processor communication this method attains good computational
speed-up and efficiency. Large systems of up to $65536^2$ spins have been
simulated at updating speeds of $11$ nanosecs/site ($90.7 \times 10^6$ spin
updates/sec) using state-of-the-art supercomputers. In the second part of
the article we use the cluster algorithm to study the relaxation of
magnetization and energy on large Ising models using Swendsen-Wang
dynamics. We found evidence that exponential and power law factors are
present in the relaxation process as has been proposed by Hackl {\it et
al}.  The variation of the power-law exponent $\lambda_M$ taken at face
value indicates that the value of $z_M$ falls in the interval $0.31 - 0.49$
for the time interval analysed and appears to vanish asymptotically.

% {\it Keywords: Ising Model, Cluster Labeling, Percolation, Monte Carlo
% Simulations, Parallel Algorithms, Accelerated Dynamics, Relaxation Dynamics.}

\bigskip

\bigskip

* Current Address: GSIA, Carnegie-Mellon University, Pittsburgh, PA 15213.

\newpage

\baselineskip=16pt

\bigskip
{\Large \bf 1. Introduction.}
\medskip

Over the last twenty years Monte Carlo simulations have become an important
and reliable calculation method in Statistical Mechanics and Field
% Theory\cite{Rebbi83,Binder86,Stauffer92,Stauffer93}. Large-scale
Theory [1-4]. Large-scale
high-resolution Monte Carlo simulations have provided means to compute
critical exponents, transition temperatures, cumulants and other
critical properties with
unprecedented
%%accuracy\cite{Rapaport85,Ferrenberg91,Baillie92,Burkitt89,Baillie91,Stauffer91,Kertesz92,Flanigan92,Hackl93,Bauernfeind94,Hackl94,Barkema94}.
accuracy[5-20].

Efficient cluster labeling algorithms are needed for large-scale
simulations of spin models using cluster
dynamics\cite{Swendsen87,acceldyn} and other physical systems such as
percolation clusters, nucleation droplets, polymers, fractal structures,
and particle tracks.  Cluster labeling is related to the problem of finding
the connected components of a graph, which has applications in computer
vision, image processing and network analysis, among others.

Many general parallel algorithms for finding connected components have been
introduced in the Computer Science
%literature\cite{Cormen73,Horowitz78,Shiloach82,Vishkin84,Quinn84,Gopalakrishnan85,Lim86,Tucker86,Embrechts89,Woo89,Cypher89,Blelloch90,Bader94,Greiner93,Choudhary94}.
literature [23-37]. Most of these methods are designed for SIMD
architectures using high-level language algorithmic descriptions. These
algorithms can be used for lattice problems, such as in the case of image
processing, but are not well suited for large-scale Monte Carlo simulations
which require the repeated labeling of complicated structures (fractals).
The main problem with these general methods is their absolute performance
when implemented in real parallel computers.

 Over the last five years a number of SPMD (Single Program Multiple Data)
parallel cluster labeling methods for Monte Carlo simulations have been
%%introduced\cite{Burkitt89,Baillie91,Kertesz92,Flanigan92,Hackl93,Bauernfeind94,Hackl94,Barkema94}.
introduced [8, 10, 15-20].
These methods have attained scalability with different degrees of success.
A few of these methods are scalable and can be used with large numbers of
processors\cite{Baillie91,Flanigan92,Bauernfeind94,Hackl94}. SIMD and
vector cluster labeling algorithms explicitly designed for Monte Carlo
simulations have also been
% introduced\cite{Baillie91,Brower91,Apostolakis92,Coddington92,Mino91}
introduced [10-13, 38]
but
their absolute performance still lags behind the one attained by the best
SPMD algorithms.

In this paper we present an optimized version of a SPMD relaxation-based
algorithm introduced in ref.~\cite{Flanigan92}.  The algorithm is general
and can be applied to higher dimensional systems.  We will concentrate on
the problem of cluster labeling for the $2D$ Ising Model with Swendsen-Wang
dynamics\cite{Swendsen87}.  A percolation process\cite{Stauffer91} is used
to define bonds between aligned spins.  The bonds are thrown with
probability $p_{bond} = 1 - e^{-2\beta}$, and the clusters of connected
spins, the Coniglio-Klein\cite{Coniglio80,Coniglio89} percolation clusters, are
flipped with 50\% probability.  At the critical point the clusters span the
system and labeling information has to propagate across the entire
computational domain.  The basic algorithm is discussed in
ref.~\cite{Flanigan92} and here we will give a brief review of the improved
version and present new data.  In the second part of the paper we will show
new results of the application of the algorithm to study the relaxation of
large $2D$ Ising models.

\bigskip
{\Large \bf 2. Description of the Algorithm.}
\medskip

Physical space is divided into rectangular cells in such way that each cell
is assigned to one processor (see Fig. 1). The algorithm labels the
clusters in two
stages: first it finds all the clusters inside each processor using a
serial algorithm, and then it performs a global relaxation process where
processors exchange clusters labels with nearest neighbors until a fixed
point is reached.  The operations of the algorithm are shown in Figs. 2 and
3. The Cluster-Labeling procedure can be described as follows:

\bigskip

{\bf Procedure Spin-Dynamics \& Cluster-Labeling:}

\begin{itemize}

\item[(i)] Define connectivity for the sites (spins) by
throwing the Swendsen-Wang percolation bonds.

\item[(ii)] Apply a serial algorithm (see procedure Local-Labeling below)
 to label the
clusters inside each processor independently.  At boundary sites the
off-node bonds are ignored for now. At the end all sites are labeled with
their ``local root'' labels which are then globalized (i.e. made unique
over the whole system).

\item[(iii)] Iterate several relaxation cycles  (see procedure
Relaxation below), exchanging local root labels with neighboring processors
until all processors detect no change in the labels. At the end all sites
get their final global label from their local roots.

\item[(iv)] Clusters of spins are flipped with 50\% probability
and measurements of
relevant quantities are accumulated (energy, magnetization etc.).

\end{itemize}

The serial local algorithm we used in ref.~\cite{Flanigan92}, and which is
based on the Hoshen-Kopelman\cite{Hoshen76} algorithm, was somewhat
inefficient in terms of cache memory utilization and other
implementation-dependent characteristics. To improve it we explored
different variations of it until we found one that is significantly better.
This particular serial algorithm can be described as follows.

\bigskip

{\bf Procedure Local-Labeling:}
\medskip
( cluster labels are used as ``pointers'' to local sites.)
\medskip
\begin{itemize}

\item[(i)] All sites are initialized with a unique
label (i.e. each site points to itself).

\item[(ii)] For each site do the following, starting from the north-west
corner and moving down row-wise:

\begin{itemize}

\item[(a)] If the neighbor to the north is connected then
follow the label pointer until the root is found (site pointing to itself).
If the root label is less than the current label then set the current label
(and current root) to the root label.

\item[(a)] If the neighbor to the east is connected set its label to the
current
label.

\end{itemize}

\item[(iii)] After this has been done for all nodes then do a final collapse of
the connectivity trees, defined by the previous operations, by making every
site point to the root label of the cluster it belongs to.

\end{itemize}

The connections across processor boundaries are ignored during the local
labeling. Once the local labeling has been completed (see Fig. 2) a number
of relaxation cycles are iterated in order to label the clusters globally
(across processors).  Each cycle consists of interchanging, comparing and
resetting labels with neighboring processors.  The relaxation procedure
consists of the following:

\bigskip

{\bf Procedure Relaxation:}

\begin{itemize}

\item[(i)] A preparation step is executed to set up a list of pointers
to the local roots on each node boundary (see last panel in Fig. 2).

\item[(ii)] A number of relaxation steps are iterated. To do
 this each processor interchanges boundary labels with the neighbors in
each direction for those boundary sites which have connections off
processor.  The labels are sent in a single block of data using standard
message-passing calls.  The local root labels are compared with the ones
received from the neighbors and then set to the minimum values.  This is
done for the four directions (north, east, south, west) before checking for
the termination condition (all nodes detect no change in the labels).

\end{itemize}

We found that this algorithm is well suited for Monte Carlo simulations and
that even for large numbers of processors the time spent on the relaxation
cycles is relatively small. The relaxation procedure appears to be a wasteful
operation at first sight; however, the algorithm is efficient because
communications are done by exchanging large blocks and the label comparison
and resetting overhead is kept to a minimum. The connectivity of the
``renormalized'' processor-grid lattice is much simpler than that of the
original system and therefore the relaxation process converges very
quickly. Total execution times are dominated by local labeling not by
communications. All these results in good scaling and performance.  As we
will see in the next section this algorithm can attain updating speeds of
about $90$ million spin-updates per second, on a 256-node CM-5E, and is
considerably faster than comparable algorithms on vector
supercomputers\cite{Wolff89,Mino91}. Surprisingly enough it is even faster
than some of the high-performance single-spin flip algorithms using
multi-spin coding\cite{Munkel93}.

\bigskip
{\Large \bf 2. Scaling and Numerical Results.}
\medskip

Taking into account the fact that the cluster labeling algorithm
 operates at the
core of equilibrium Monte Carlo simulations one has to worry more about
average than worst case performance. The probability of obtaining a given
configuration of clusters is determined by the Boltzmann weight of that
configuration.  The probability of observing the worst case is
negligible. On the other hand, ``typical'' configurations often contain
fractal structures which are relatively hard to label.

In order to make a meaningful scaling analysis one has to concentrate on
average execution times defined over critical equilibrium Monte Carlo
configurations.  A detailed complexity and scaling analysis was presented
in ref.~\cite{Flanigan92}.  Here we will review the main results.

The  total  time  to  perform  the   cluster  labeling  consists   of   two
contributions: a  ``local  time'', the time  spent by the  serial algorithm
inside  each processor, and a ``relax  time'' which is the time spent
in the global relaxation procedure until completion,

\begin{equation}
T_{parallel} \; = \; T_{local} + T_{relax}\; = \; a n + b p^{d_{min}/2}
n^{1/2},
\end{equation}
\noindent
where $n = l \times l$ is the size for the subsystem assigned to each
processor, $a$ and $b$ are constants that characterize the computation and
communication rates for a given machine, and $p$ is the total number of
processors. The scaling of the local part is assumed to be basically
$O(n)$. In principle it is $O(n \log^* n)$ but, as is discussed it ref.
\cite{Horowitz78}, $\log^* n$ can be safely considered a constant.
The exponent $d_{min}$ is the {\it chemical distance} exponent that plays
the role of a dynamical critical exponent for the labeling
process\cite{Herrmann88,Miranda91,Flanigan92}.  If $n$ is large compared
with $p$, and the communication to computation ratio $a/b$ is small, the
scaling will be dominated by the local part.  This can be better seen by
analyzing the speed-up function $S(n, p)$, which is the ratio between the
serial and parallel times,

\begin{equation}
S(n, p) \; = \; { T_{serial} \over T_{parallel}} \;
= \; { anp \over an + b p^{d_{min}/2} n^{1/2}}.
\end{equation}
The speed-up improves with large $n$ and gets worse as $p$
increases. The speed-up as a function of $p$ for fixed
total system size $N$ is given by,

\begin{equation}
S_N(p) \; = \; { aNp \over aN + b p^{(d_{min} + 1)/2} N^{1/2}},
\end{equation}
\noindent
and the corresponding efficiency $E_N(p) = S_N(p)/p$ is,

\begin{equation}
E_N(p) \; = \; { 1 \over 1 + {b \over a} [p^{(d_{min} + 1)} N^{-1}]^{1/2}}.
\end{equation}
The efficiency decreases as the number of processors increases
because inter-processor communication times eventually dominate.  In
practice, it is important to make $b/a$ as small as possible to reduce
absolute times. The efficiency is a universal function of $p^{(d_{min}+1)}
N^{-1}$ and therefore the only important parameter in the simulation, from
a computational point of view, is the value of $p^{(d_{min}+1)}N^{-1}$.
The larger $N$ is in relation to $p$, the better the algorithm will perform.

Our implementation of the algorithm was done using standard C language plus
message-passing calls (CMMD library) on the CM-5E supercomputer. We
measured execution times for different total system sizes $N$ and numbers
of processors $p$ for 2D Ising Swendsen-Wang simulations. We obtained:
$a=2.79 \times 10^{-6}$ secs. and $b = 6.2 \times 10^{-6}$ secs.,
and therefore $b/a =
2.22$. The timings are shown in Table I.

Fig. 4 shows local, relax, and total times for different system sizes on a
64-node CM-5E.  Measurements times (energy, magnetization, etc.)  were not
included.  As we can see in the figure, local times dominate except for
small $N$.  Typical updating times are of the order of $40$ nanosecs/site
and $10$ nanosecs/site for $64$ and $256$ node CM-5E machines respectively.

\bigskip
Table I: Timings for 2D Swendsen-Wang dynamics on CM-5E supercomputers.
\smallskip

\begin{center}
{\small
\begin{tabular}{|l|l|l|l|l|l|l|} \hline
System Size & number of procs. & Local time & Relax time & Total time &  Speed
&   Time/site  \\
    $N$     &   $p$        & [secs]     &   [secs]   &  [secs]   &  [$10^6$spin
updates/sec] & [nanosecs] \\ \hline
128$^2$     & 32           & 0.0012     &  0.0014    & 0.0026     & 6.3 & 164
\\
256$^2$     & 32           &  0.005     &  0.002     & 0.007      & 9.4 & 104
\\
512$^2$     & 32           &  0.020     &  0.003     & 0.023     & 11.3 & 88.3
\\
1024$^2$    & 32           &  0.078     &  0.007     & 0.085      & 12.3 & 81.1
\\
2048$^2$    & 32           &  0.312     &  0.017     & 0.329      & 12.7  &
78.5 \\
4096$^2$    & 32           &  1.24      &  0.04      & 1.28       & 13.1  &
76.5 \\
8192$^2$    & 32           &  5.15      &  0.08      & 5.23       & 12.8  &
78.0 \\
16384$^2$   & 32           & 22.19      &  0.16      &  22.4      & 12.0  &
83.3 \\ \hline
256$^2$     & 64           & 0.002      &  0.002     &  0.004     & 15.2  &
66.4 \\
512$^2$     & 64           & 0.010      &  0.003     &  0.013     & 20.2  &
48.2 \\
1024$^2$    & 64           & 0.039      &  0.006     &  0.045     & 23.3  &
42.5 \\
2048$^2$    & 64           & 0.157      &  0.013     &  0.169     & 24.8  &
40.4 \\
4096$^2$    & 64           & 0.623      &  0.027     &  0.650     & 25.8  &
38.7 \\
8192$^2$    & 64           & 2.580      &  0.060     &  2.640     & 25.4  &
39.3 \\
16384$^2$   & 64           & 11.12      &  0.12      &  11.24     & 23.9  &
41.9 \\
32768$^2$   & 64           & 46.27      &  0.24      &  46.52     & 23.1  &
43.3 \\  \hline
512$^2$     & 256          & 0.002      &  0.004     &  0.006     & 41.6  &
24.02 \\
1024$^2$    & 256          & 0.009      &  0.006     &  0.015     & 69.4  &
14.41 \\
2048$^2$    & 256          & 0.037      &  0.011     &  0.048     & 87.2  &
11.47 \\
4096$^2$    & 256          & 0.149      &  0.021     &  0.170     & 98.5  &
10.15 \\
8192$^2$    & 256          & 0.596      &  0.049     &  0.645     & 104   &
9.61 \\
16384$^2$   & 256          & 2.430      &  0.096     &  2.520     & 106   &
9.40 \\
32768$^2$   & 256          & 10.63      &  0.105     &  10.73     & 100   &
9.99 \\
65536$^2$   & 256          & 47.10      &  0.261     &  47.36     & 90.7  &
11.0 \\ \hline
\end{tabular}
}
\end{center}

For very large systems the performance decreases slightly due to decreased
cache memory usage. This effect is not taken into account in the scaling
model. Simulating a $32768^2$ ($65536^2$) system requires about $84$ Mbytes
of local memory on a $64$ ($256$) node CM-5E.  This is about 5 bytes per
site: 4 bytes (one integer) for the label and one byte for the spin.  The
scaling behavior of the model agrees well with the actual measured times.
The local times scale linearly with $n$, and the relaxation times with
$n^{1/2}$ as expected.  The speed-up $S_N(p)$, as a function of the number
of processors for different system sizes, is shown in Fig. 5.  Notice that
even for small system sizes, such as $L=512$, the speed-up increases
without saturation even for $p$ equal to $256$ processing nodes.  The
$32768^2$ and $65536^2$ lattices are simulated with more than 99\%
efficiency on 64 and 256 processors respectively.  The efficiency $E_N(p) =
S_N(p)/p$ as a function of $p$ is shown in Fig.  5.  When these points are
plotted as a function of $[p^{(d_{min}+1)}N^{-1}]^{1/2}$, as shown in Fig.
6, they show the universal scaling behavior predicted by Eq.  4: data for
different system sizes and number of processors collapses reasonably well
to a single curve.

 The relaxation method can be extended in a hierarchical style, for example
using the multi-grid approach introduced in ref.~\cite{Brower91}, but in
practice this is unnecessary because nearest neighbor relaxation is a very
efficient operation for global labeling even on large processor grids.

\bigskip
{\bf  3. Relaxation of Large 2D Ising Model with Swendsen-Wang Dynamics.}
\medskip

In this section we present results of a study of the relaxation of
magnetization and energy on large 2D Swendsen-Wang Ising models. The
typical simulation consists of setting all spins up and then letting the
system relax to equilibrium at $T_c$.  Several relaxation experiments are
performed to obtain significant statistics.  When enough data is available
different scaling models can be tested and the value of $z$ estimated.
This method has been used for local Ising dynamics in two and three
dimensions by Stauffer\cite{Stauffer91}, Stauffer and
Kertesz\cite{Stauffer91-2} and Miranda\cite{Miranda91}.  Stauffer and
Kertesz\cite{Kertesz92} simulated large Swendsen-Wang Ising systems with up
to $6400^2$ spins and found large corrections to scaling and finite size
effects.  They also found that the time dependent exponent $z(t)$ vanishes
linearly in $1/t$ implying $z=0$ in agreement with previous results of
Heermann and Burkitt\cite{Burkitt89}, and others.  Tamayo\cite{Tamayo93}
performed a similar analysis with systems up to $32768^2$ spins and found
relaxation behavior for intermediate and large times which was consistent
with a power law decay and $z= 0.25 \pm 0.05$.  Later, Hackl {\it et al}
\cite{Hackl94} studied systems with up to $17920^2$ spins and proposed a
 relaxation ansatz
combining power law and exponential behavior. They found that this ansatz
was able to fit the data reasonably well and explained the difference in
the behaviors observed by Stauffer and Kertesz\cite{Kertesz92} and
Tamayo\cite{Tamayo93}.  Their power law exponents correspond to $z_M =$
0.43 and $z_E =$ 044.  These values are higher than the value $z_M=$ 0.25
found by Tamayo, however a comparison of Hackl {\it et al}'s raw
relaxation data with Tamayo's showed good agreement\cite{Hackl94-2}.

In order to better understand the nature of these differences and explore
the complex relaxation of energy and magnetization we performed relaxation
experiments with large $2D$ systems with $L=2048$, $L=16384$ and $L=32768$
spins.  Each experiment consisted of setting all spins up and letting the
system run for 100 Swendsen-Wang time steps with the temperature set to
$T_c$. In total we performed 5,000 relaxation experiments for $L=2048$, 210
for $L=16384$ and 46 for $L=32768$. Our analysis will focus on the data for
the $L=32768$ system.

Figs. 8 and 9 show the average magnetization and energy as a function of
time for different lattices. For short times the relaxation is basically
the same for all lattice sizes but for longer times is different due to
finite size effects. The behavior at short times appears to be dominated by
an exponential factor as was observed by Kertesz and
Stauffer\cite{Kertesz92}.  At intermediate and longer times the behavior
appears to be dominated by a power law\cite{Tamayo93}. To account for this
exponential/power-law phenomenology Hackl {\it et al} proposed an ansatz
combining exponential and power law decay\cite{Hackl94},

\begin{equation}
M(t) \sim (t + \Delta_M)^{- \lambda_M} e^{-b_M t}
\end{equation}
\noindent
and similarly for the energy,

\begin{equation}
E(t)  \sim  (t + \Delta_E)^{- \lambda_E} e^{-b_E t} .
\end{equation}

If one considers an overall amplitude in addition to $b_M$, $\lambda_M$ and
$\Delta_M$ the model has 4 parameters. In order to fit this model to Monte
Carlo data Hackl {\it et al} found useful to define the following auxiliary
functions,

\begin{equation}
h_M(t) = - {1 \over M(t)} {dM(t) \over dt}
\end{equation}

and

\begin{equation}
g_M(t) = - {1 \over h_M(t+2) - h_M(2)} .
\end{equation}

These functions can be expressed in terms of $\Delta_M$,
$b_M$ and $\lambda_M$ as follows,

\begin{eqnarray}
h_M(t) & = & {\lambda_M \over t + \Delta_M} + b_M\\
\label{fit}
g_M(t) & = & {1 \over \lambda_M}
   \left[ {(\Delta_M + 2)^2 \over t} + \Delta_M + 2 \right] .
\end{eqnarray}

Analogous equations can be defined for the energy. The reason for this
reformulation is that $g_M$ ($g_E(t)$) is independent of $b_M$ ($b_E$) and
then a standard linear fit can be done for $g_M$ ($g_E$) {\it vs} $1/t$.
In this linear fit the slope and intersect correspond to $1/\lambda_M
(\Delta_M + 2)^2$ and $1/\lambda_M (2 + \Delta_M)$ respectively. Hackl {\it
et al} fitted this model to their data and found the following values for
the parameters\cite{Hackl94}: $b_M = 0.0043 \pm 0.0010$, $\lambda_M = 0.29
\pm 0.03$, and $\Delta_M = 4.5 \pm 0.6$ for the magnetization and $b_E =
0.0281 \pm 0.0020$, $\lambda_E = 2.26 \pm 0.08$, and $\Delta_E = 5.9 \pm
0.25$ for the energy.

The exponent $b_M$ ($b_E$) was obtained using the values of $\lambda_M$
($\lambda_E$) and  $\Delta_M$ ($\Delta_E$) to calculate the function,

\begin{equation} \label{beq}
\tilde{b}_M(t) = \lambda_M \log \left[{ t + 1 + \Delta_M \over t +
\Delta_M}\right] +  \log \left[{M(t+1) \over M(t)} \right] ,
\end{equation}

\noindent
and averaging $\tilde{b}_M(t)$ ($\tilde{b}_E(t)$) over long times where it
is roughly constant. Another way in which they obtained $b_M$ ($b_E$) is by
imposing the condition $M(t=0) = 0$ and $E(t=0) = -2$ on the ansatz,

\begin{equation}
b_M(t) = - {1 \over t} \left[ \log {M(t) \over M(0)}
        + \lambda_M \log{t + \Delta_M \over \Delta_M} \right] ,
\end{equation}
\noindent
and then averaging the roughly constant $b_M(t)$ ($b_E(t)$) for
 intermediate and long
times.

  The exponential/power-law ansatz fitted their data reasonably well. From
the values of $\lambda_M$ and $\lambda_E$ they also computed
values for $z$ using the following relations,

\begin{eqnarray}
\lambda_M & = & \beta / \nu z\\
\lambda_E & = & (1 - \alpha)/ z,
\end{eqnarray}
\noindent
which produced $z_M$ = 0.43 and $z_E$ = 0.44. They did not find evidence of
a logarithmic factor in the energy.

In our study we followed a similar methodology. We analyzed data from 46
relaxation experiments with the $32768^2$ system. Figs. 8 and 9 show the raw
relaxation data for energy and magnetization on three different lattice
sizes. From these plots we obtained instantaneous slopes and values of
$z_M(t)$ and $z_E(t)$ as a function of time (see Figs. 10 and 11). These
values of $z_M(t)$ and $z_E(t)$ appear to scale roughly linearly with $1/t$
implying $z \neq 0$, as was found in ref.
\cite{Tamayo93}, but the oscillating behavior near $t=0$ makes difficult
to make an accurate extrapolation for $1/t \rightarrow 0$ and compute an
effective $z$. If the decay process were a pure power-law then according to
the figures the values of $z_M$ and $z_E$ will be in the range $0.2 -
0.3$.  In addition to the instantaneous slopes we also computed $g_M$
($g_E$) which are shown in Figs. 12 and 13.  A linear fit of these data
using Eq.
\ref{fit}, from $t=1$ to $t=80$, produced the following values,

\begin{eqnarray}
b_M & = & 0.0043 \pm 0.0005      \\
\lambda_M & = & 0.28 \pm 0.01 \\
\Delta_M & = & 5.2 \pm 0.1,
\end{eqnarray}

and

\begin{eqnarray}
b_E & = &   0.031 \pm 0.001 \\
\lambda_E & = &  2.2 \pm 0.1  \\
\Delta_E & = & 7.5 \pm 0.1.
\end{eqnarray}

These numbers agree well with the values of Hackl {\it et al}\cite{Hackl94}
for all the parameters except for $\Delta_E$.  The exponent $b_M$ ($b_E$)
was computed by evaluating Eq. \ref{beq} using the values of $\lambda_M$
($\lambda_E$) and $\Delta_M$ ($\Delta_E$) from the $g_M$ ($g_E$) fit and
averaging $\tilde{b}_M$ ($\tilde{b}_E$) from $t=10$ to $t=60$. This is
shown in Figs.  14 and 15.

 In Figs. 16 and 17 we have plotted the original data and the fit described
by the ansatz.  The power law/exponential ansatz,	 parameterized
either with the values obtained in this paper or with Hackl {\it et al}
values, describes the data reasonably well for initial and intermediate
times although it appears to slightly underestimate the values for long
times. To study this effect we repeated the analysis but now fitted the
data from $t=10$ to $t=80$ in this way giving more weight to long time
behavior. The fit produced somewhat different values for the magnetization
parameters,

\begin{eqnarray}
b_M & = &  0.0020 \pm  0.0005 \\
\lambda_M & = & 0.38 \pm 0.01 \\
\Delta_M & = & 7.0 \pm 0.1
\end{eqnarray}
\indent
but basically identical ones for the energy,

\begin{eqnarray}
b_E & = &  0.031 \pm 0.001            \\
\lambda_E & = &  2.2 \pm 0.1  \\
\Delta_E & = & 7.5 \pm 0.1
\end{eqnarray}

 In this case the ansatz for the magnetization fits better the data for
long times as can be seen in Fig. 16. In fact, the $\lambda_M = 0.38$
implies $z_M$ = 0.33 which is now closer to the value Tamayo found fitting
a single power law decay. To expose this apparent different behavior of the
short and long time magnetization data we performed an incremental fit of
$g_M$ ($g_E$) {\it vs} $1/t$ always starting at $t=1$ but ending at
multiples of $10$; the results are,

\begin{center}
{\small
\begin{tabular}{|ll|ll|ll|} \hline
initial time & final time & $\Delta_M$ & $\lambda_M$ & $\Delta_E$ & $\lambda_E$
\\ \hline
    1       &     11     &   4.70    &   0.253   &    7.58 &     2.22 \\
    1       &     21     &   4.94    &   0.267   &    7.52 &     2.20 \\
    1       &     31     &   5.04    &   0.273   &    7.52 &     2.20 \\
    1       &     41     &   5.11    &   0.278   &    7.52 &     2.20 \\
    1       &     51     &   5.18    &   0.280   &    7.52 &     2.20 \\
    1       &     61     &   5.20    &   0.283   &    7.52 &     2.20 \\
    1       &     71     &   5.22    &   0.284   &    7.52 &     2.20 \\
    1       &     81     &   5.29    &   0.288   &    7.52 &     2.20 \\
    1       &     91     &   5.33    &   0.291   &    7.52 &     2.20 \\ \hline
\end{tabular}
}
\end{center}

Compare that with a similar incremental fit but now starting from $t=10$,

\begin{center}
{\small
\begin{tabular}{|ll|ll|ll|} \hline
initial time & final time & $\Delta_M$ & $\lambda_M$   & $\Delta_E$ &
$\lambda_E$ \\ \hline
    10       &     20     &  5.33      & 0.292       &   7.48  &   2.18 \\
    10       &     30     &  5.63      & 0.305	     &   7.53  &   2.19 \\
    10       &     40     &  5.96      & 0.323	     &   7.54  &   2.20 \\
    10       &     50     &  6.18      & 0.335	     &   7.54  &   2.20 \\
    10       &     60     &  6.49      & 0.352	     &   7.55  &   2.20 \\
    10       &     70     &  6.58      & 0.357	     &   7.55  &   2.20 \\
    10       &     80     &  6.98      & 0.381	     &   7.55  &   2.20 \\
    10       &     90     &  7.34      & 0.402       &   7.55  &   2.20 \\
\hline
\end{tabular}
}
\end{center}

There are several conclusions we draw from this. The first is that the
ansatz describes the energy relaxation very well. The energy results are
consistent and very robust independent of the particular subset of data
being used for the fit. For the magnetization the ansatz works well for
short or intermediate times but for long times there is a trend by which
the ansatz underestimates the data. This manifests as the change of values
of $\Delta_M$ and $\lambda_M$ as one adds longer time data points to the
fit.  Assuming this trend is significant then the behavior of the
magnetization relaxation at long times might be more complex than expected.
The variation of $\lambda_M$ taken at face value indicates that the value
of $z_M$ falls in the interval $0.31 - 0.49$ for the time interval
analysed and appears to vanish asymptotically.
For the energy $z_E$ appears to be
around $0.45$ and we didn't find evidence of a logarithmic factor.

 In summary, we found evidence that exponential as well as power law
factors are present in the relaxation of the energy and magnetization of
large $2D$ Ising models with Swendsen-Wang dynamics. The ansatz proposed by
Hackl {\it et al}, which combines an exponential factor with a power law,
describes the energy relaxation very well. The magnetization relaxation is
also described well for short and intermediate times but the behavior
appears to be more complex at longer times.

  The effect of different initial conditions (eg. $M(t=0) \ne 1$), as has been
studied in refs. \cite{Colonna-Romano94,Janssen88}, was not considered here
but will be the subject of future work.

\bigskip
{\bf  Acknowledgments.}
\medskip

We want to thank L. Colonna-Romano, A. I. Mel\'cuk, H. Gould, W. Klein, L.
Tucker, R. Hackl and D.  Stauffer, for comments and correspondence and J.
Mesirov and B. Lordi of Thinking Machines Corp.  for supporting this
project.  We wish to acknowledge the DOE Advanced Computing Lab.  at Los
Alamos, the Naval Research Laboratory and Thinking Machines Corp.  for
providing the computer time that was needed to perform the simulations
reported in this work.

\end{document}